\documentclass[11pt,dvips]{article}
\usepackage{epsfig,times} 
\usepackage{picinpar}
\usepackage{wrapfig}
\usepackage{floatflt}
\makeatletter
\renewcommand{\section}{\@startsection{section}{1}{0in}
	{0.4\baselineskip}{0.1\baselineskip}{\Large\bf}}
\renewcommand{\subsection}{\@startsection{subsection}{2}{0in}
	{0.25\baselineskip}{-\baselineskip}{\large\bf}}
\renewcommand{\subsubsection}{\@startsection{subsubsection}{3}{0in}
	{0.1\baselineskip}{-\baselineskip}{\normalsize\bf}}
\makeatother

\begin{document}
%
\pagestyle{myheadings}
\markright{ To be presented at $26^{\rm th}$ International Cosmic Ray Conference Salt Lake City - Utah 1999 - HE 2.5.06}
\begin{center}
{\LARGE \bf A method to estimate primary composition with the Auger Observatory}
\end{center}
\begin{center}
{\bf L.A. de Carvalho$^{1}$ and C. Dobrigkeit$^{1}$}\\
{\it $^{1}$Departamento de Raios C\'osmicos e Cronologia, Instituto de F\'{\i}sica ``Gleb Wataghin'' \\ Universidade Estadual de Campinas, C.P 6165, Campinas, SP, CEP 13083-970, Brasil}
\end{center}
\begin{center}
{\large \bf Abstract\\}
\end{center}
\vspace{-0.5ex}
We explore the feasibility of estimating primary cosmic ray composition at ultra high energies from the sum of muon, electron and photon densities and the depth of maximum of extensive air showers detected by the Auger Observatory. From the information of the Fluorescence ($X_{max}$) and water Cerenkov detectors ($\rho(1000)$)  we infere the most probable type of primary which originated the shower. The method is tested simulating Extensive Air Showers (EAS) at energies up to $10^{20}$ eV using a version of MOCCA/Sibyll adapted for Dec Alpha Servers. We also discuss some results on the mass composition using the depth of the shower maximum and the lateral distribution of showers in the energy range $10^{18}$-$10^{20}$ eV.
\vspace{1ex}
%
\section{Introduction:}
\label{intro.sec}

In recent years ultra-high-energy cosmic rays have received renewed attention from the whole physics community. Most of the interest for this subject comes from the observation of  eight extremely energetic particles (Auger Collaboration, 1997),  inspite of the so-called GZK cutoff.  In 1966 Greisen and, independently, Zatsepin and Kuzmin predicted the suppression of cosmic rays of energies as high as $10^{20} eV$ or above, an effect known as the GZK cutoff. As they pointed out, this suppression is simply due to the fact that protons with energy above about $4.1 \times 10^{19} eV$ should interact with the cosmic microwave background radiation, implying that no such high-energetic particles should be observed.

Since there had been observations of those air showers that seem to have been induced by cosmic rays of that high energies, the intriguing challenge is that possible sources of these particles should not be located far away, otherwise they would have been suppressed by the GZK effect. The details about those events can be found in the Design Report of the Auger Project (1997).

In order to observe air showers induced by UHE cosmic rays, the Auger Collaboration has chosen to combine two different detection techniques already explored by two long-lasting collaborations. These techniques involve the observation of the Cerenkov light emitted by ultra-relativistic electrons and muons in water tanks, as used in the experiment of Haverah Park (Lawrence, Reid \& Watson 1991), and the observation of the fluorescent light with mirrors and photomultipliers, as done in the Fly's Eye experiment (Baltrusaitis et al 1985). This fluorescent light is emitted by $N_{2}$ molecules in the atmosphere when excited by air shower particles, mainly electrons and positrons.

Measurements  of  arrival times give an estimate the shower direction, also assumed as that of the primary. From this information and the signals in the water tanks and/or fluorescent detectors, the fraction of the primary energy gone into electromagnetic particles and the depth of the shower maximum can be inferred. Those  combined observations will give an estimate of the primary mass composition and also a hint on the possible origin of those extremely energetic particles.

Unfortunately the measurements of both Haverah Park and Fly's Eye experiments do not allow to estimate the mass of the primary at an event-to-event basis, but only a mean composition for a given energy range. Sokolsky et al (1992) presented a detailed report on this subject, to serve as a reference guide about the detection techniques, their experimental results and methods to characterize UHE primary cosmic rays. From this report, the difficulties inherent to this intend are clearly seen. 

Basically the quantities which have been used to estimate primary mass composition are: 
\vspace{-0.1in}
\begin{verse}
1) The ratio between muon and electromagnetic densities;\\
2) Measurements of the depth of shower maximum ($X_{max}$) and its distribution; \\
3) The elongation rate ($ER$), or the mean rate of increase of $X_{max}$ with primary energy ($E_{0}$).
\end{verse}

\vspace{-0.1in}

We have tried to explore some correlations between those quantities for simulated air showers, in order to find out some indications for primary energy and mass.

We have used the MOCCA program  (Hillas 1981) with the Sybill model (Fletcher et al. 1994) for high energy hadronic interactions. Twelve hundred air showers were run, at a thinning level of $10^{-6}$, for two types of primaries (proton and iron nucleus), three primary energies ($10^{18}$, $10^{19}$ and $10^{20}$ $eV$) and two zenithal angles ($0^{o}$ and $30^{o}$). Previous problems in the implementation of the Landau-Pomeranchuk-Migdal effect (Migdal 1956) and in choosing the seed of the random generator in Sybill (Prike 1998) had been corrected before these runs. The version of Sybill used with MOCCA was 1.5. The electromagnetic particle energy cutoff was set as $1 MeV$.

%
%
\section{$X_{max}$, $ER$ and Fluctuations Results}
\label{fluc.sec}
\begin{figwindow}[1,r,%
{\mbox{\epsfig{file=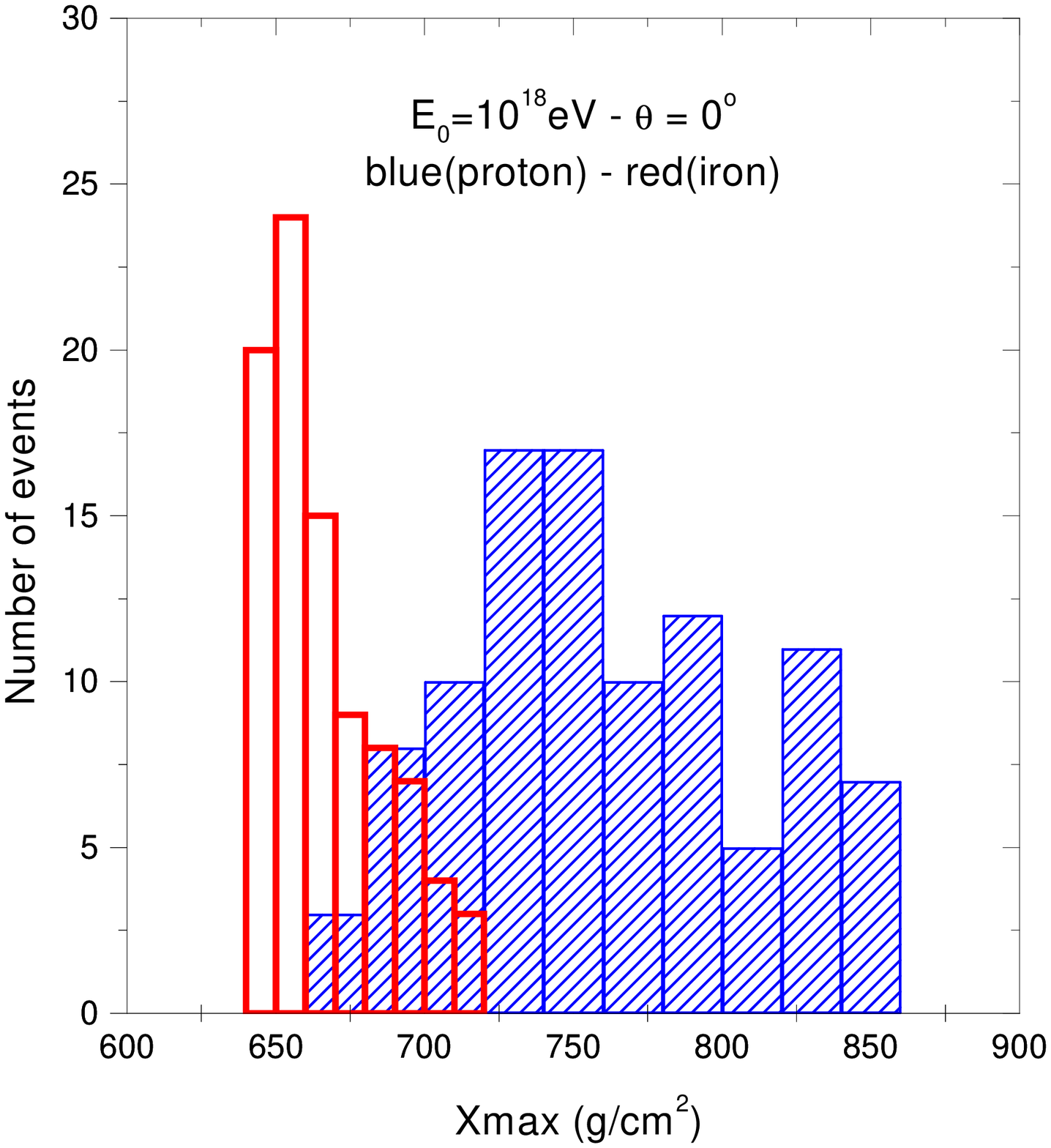,width=3.4in}}},%
{Histograms of the $X_{max}$ extracted from our simulated data.}
]
In this section we show some results obtained from those twelve hundred showers, for the three quantities cited above.

The ratio between the lateral densities of muon and electromagnetic particles is strongly dependent on the mass of the primary. As already shown  by Gaisser et al (1978), the mean ratio for iron nuclei is $60\%$ higher than for proton primaries. At the other side, the applicability of this information is very dependent on the distinction between both components in the water Cerenkov detector.

Here the superposition model is adopted for primary air nuclei interactions, and, consequently, the fluctuations in $X_{max}$ are also strongly dependent on the type of the primary particle.

Iron-induced showers show less fluctuations in the depth of maximum than proton-induced ones, as can be seen in Figure 1. It can also be observed from this histogram that in about $30\%$ of the showers it is impossible to distinguish between both types of primaries only by observing the absolute value of $X_{max}$. Of course the situation gets even worse when other primaries are included in the comparison. The observed difference between mean $X_{max}$  -values for proton- and iron-induced showers, $ER \cdot Log(A_{Fe})$, is $99\pm58 \cdot g/cm^{2}$, where the error was taken from the widths  of the gaussians fitted to the distributions. So, these fluctuations in $X_{max}$ are hard to be beaten.

Table 1 presents the values for $ER$ and for the difference between the mean $X_{max}$ for iron and proton showers, observed in this sample of twelve hundred simulated showers.

\end{figwindow}

\begin{tabular}{|cc|c||||cc|c|} \hline
Primary & $ \theta$($ ^{0}$) & $ER$ \quad ($g/cm^{2}$) & Log(Energy($eV$)) & $ \theta$($ ^{0}$) & $ER \cdot Log(A_{Fe})$ ($g/cm^{2}$) \\ \hline \hline
         &     &             & 18 &   0  & 99$\pm$58 \\ 
proton   &  0  & 63$\pm$116  & 18 &  30  & 79$\pm$69 \\
proton   & 30  & 51$\pm$100  & 19 &   0  & 78$\pm$65 \\
iron     &  0  & 61$\pm$33   & 19 &  30  & 80$\pm$48 \\  
iron     & 30  & 49$\pm$32   & 20 &   0  & 104$\pm$52 \\
         &     &             & 20 &  30  & 82$\pm$43 \\ \hline 
\end{tabular}

Sokolski et al. (1992) have concluded that the fluctuations are less sensitive to uncertainties in interaction models than absolute $X_{max}$ values themselves In their work they analyse $X_{max}$ and $ER$ of events seen by Fly's Eye and Haverah Park experiments and claim that is impossible to distinguish between two mixtures of iron and proton primaries (in one case $30\%$ protons, in the other, $80\%$ protons), although it is possible to exclude pure proton or iron composition.

%
%
\section{Our Method:}
\label{ourmeth.sec}
\begin{figwindow}[2,r,%
{\mbox{\epsfig{file=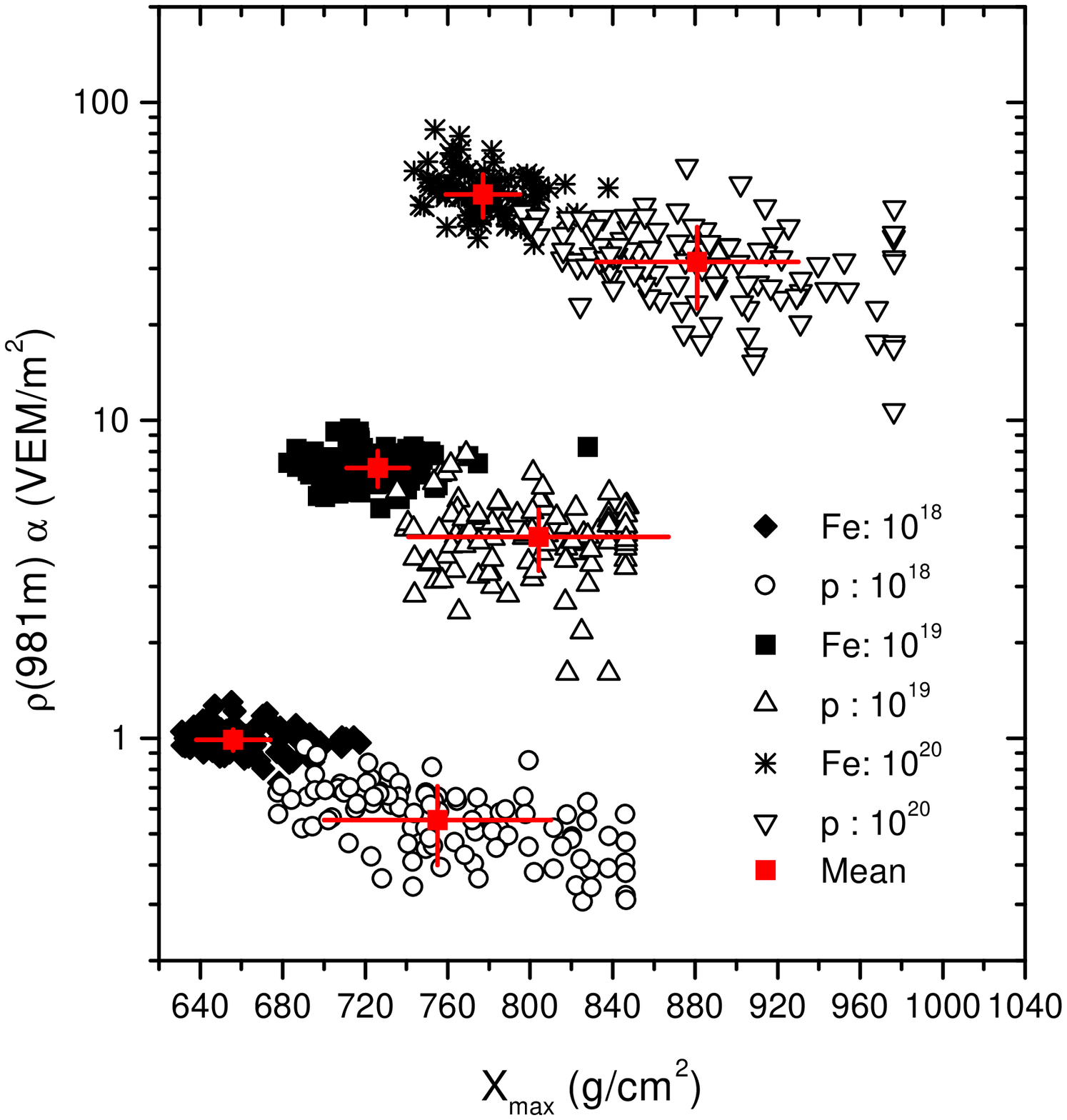,width=4.2in}}},%
{Plots of the correlation between $\rho(981)$ and its respective $X_{max}$ extracted from our simulated data.}
]
Figure 2 shows the correlation, shower by shower, between the sum of muon, electron and photon densities at $981m$ from shower core and its respective $X_{max}$ for vertical showers. Proton and iron showers are shown to separate themselves. Obviously, there are other nuclei inducing showers and a continuous plot is expected. Nevertheless, observing the mean behaviour of both quantities, it is possible to estimate, at an event to event basis, whether the primary is a light or a heavy nucleus.  

We have used the the sum of muon, electron (positron) and photon densities at $981m$ motivated by the information obtained from Haverah Park experiment that one muon with energy higher than 400 $MeV$ produces the same signal water Cerenkov tanks as an electromagnetic particle with energy higher than 250 $MeV$, being indistinguishable from it (Pryke 1996). We also remember that a photon produces a pair of electron and positron with half of its energy each, in average. So, we have used these values as energy threshold to calculate $\rho(981m)$. The calculated $\rho(981)$ will be proportional to the water Cerenkov response $VEM/m^{2}$. Similar technique was presented by Cortina et al (1997). The difference between our method and Cortina's is that they use the density of the Cerenkov light produced in the atmosphere and detected by the HEGRA AIROBICC experiment (Rhode et al 1996).
\end{figwindow}

\begin{tabular}{||c||c|c|c|c|c|c||} \hline
          & $\chi^{2}$ & $k$ & $R_{1}$ & $\beta$ & $\eta$ & $R_{0}$ \\ \hline
red line  & 0.59 & 1.39 $\times$ $10^{11}$ & 800 & 1.03 & 2.955 $\pm$ 0.004 & 4000 \\ \hline
green line& 0.17 & 1.39 $\times$ $10^{11}$ & 800 & 1.03 & 3.020 $\pm$ 0.003 & 4000 \\ \hline
\end{tabular}

In figure 2 we can also observe that $\rho(981)$ depends on the type of primary particle, $\rho(981)$ for iron showers being higher than for proton showers at same energy. We have also studied the combined lateral distribution of muons, electrons and photons by fitting the function $\rho(R) = k \cdot R^{-(\eta+\frac{R}{R_{0}})} \cdot (\frac{R}{R_{1}})^{\beta}$ to the simulated events; where $R$ is the  distance to the shower core in meter. This function was fitted to Haverah Park data (Pryke 1996) for $\theta<45^{0}$ and $10^{17}<E<5.10^{18} eV$, resulting in $\eta = 3.49 - 1.29 \cdot sec(\theta) + 0.165 \cdot Log(\frac{E}{10^{17}eV})$, $R_{0}=4000$, $R_{1}=800$ and $\beta=1.03$. In the table above we present the fitted parameters to our simulated data. The parameter $\eta$ is equal to 2.695 at same energy for the Haverah Park parametrization.

%
%
\section{Conclusions}
\label{conclu.sec}

\begin{figwindow}[1,r,%
{\mbox{\epsfig{file=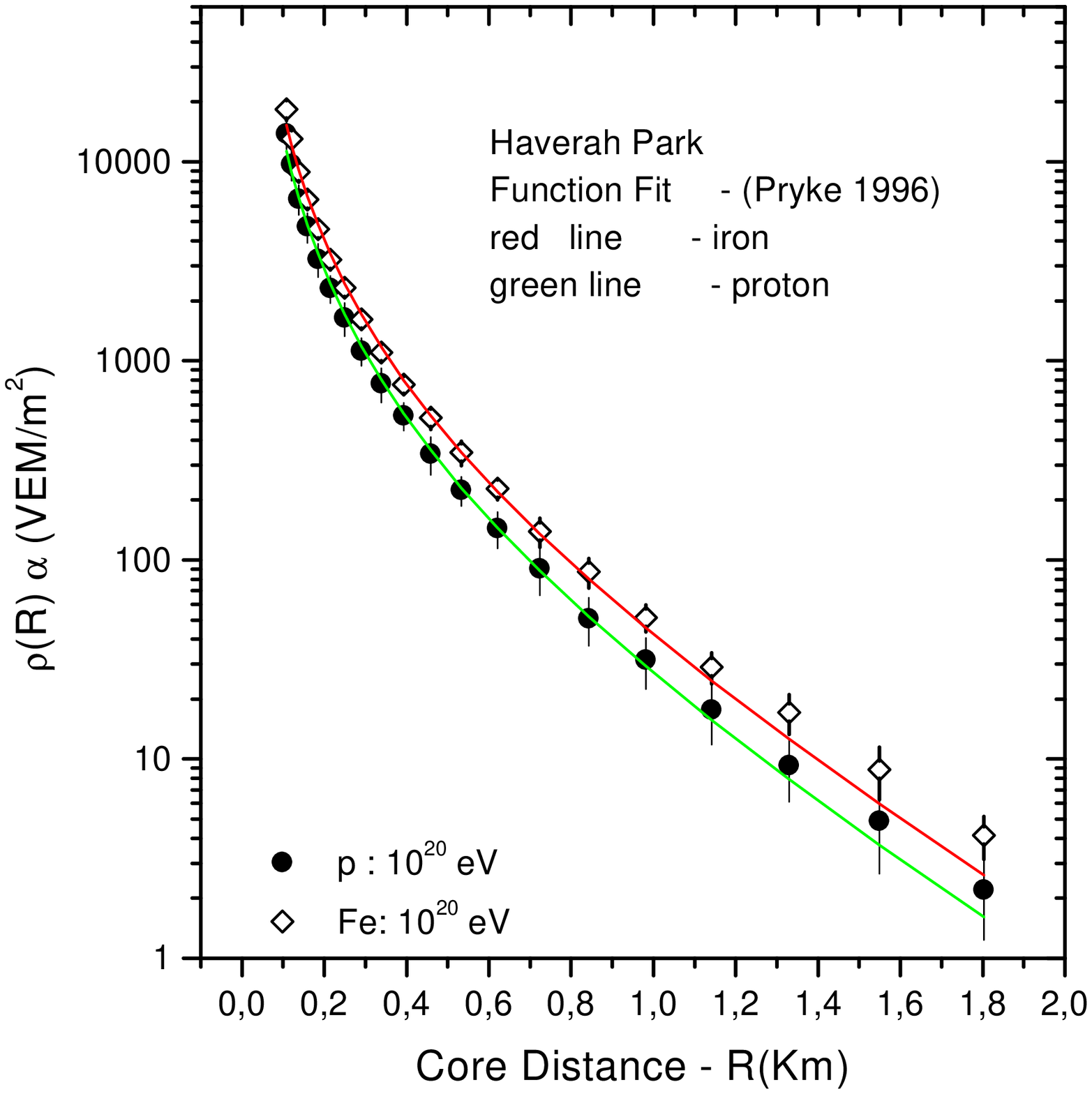,width=3.9in}}},%
{Plots of the  combined lateral distribution $\rho$ versus the distance from the shower core.}
]
We have developed a method to estimate whether the primary is a light or a heavy nucleus observing the mean behaviour of the sum of muon, electron (positron) and photon densities at $981m$ and its depth of the shower maximum. The advantage of our method is that it allows us to estimate the type of the primary particle at an event to event basis and the application of our method in the Auger Observatory data is very hopeful.

We can see that the lateral distribution for iron showers is different from that for proton showers. So, $\rho(981)$ is also different. Our lateral distribution results differ from Hillas' results (1971) that show only a slight dependence on the type of the primary particle.

We will simulate more showers for other primaries and the detection of particles by the water Cerenkov tanks to include the experimental efficiency. Finally, we would like to thank the financial support from FAPESP, A.M. Hillas and C.L. Pryke for MOCCA Code Program, and, for important help with technical problems, Mario de Castro Souza Jr.(Computing Division of IFGW  - Unicamp).
\end{figwindow}

\vspace{-.1in}
%
%
\vspace{1ex}
\begin{center}
{\Large\bf References}
\end{center}
\vspace{-.1in}
Auger Collaboration, 1997, Design Report, (Reading: http\protect{://}www\protect{.}auger\protect{.}org\protect{/}.)\\
Baltrusaitis, R.M. et al 1985, Nucl. Instrum. and Methods A 240, 410.\\
Cortina, J. et al 1997, J. Phys. G: Nucl. Part. Phys. 23, 1733.\\
%
%
Fletcher, R.S. et al 1994, Physical Review D, Vol. 50, 5710.\\
Gaisser, T.K. et al 1978, Rev. Mod. Phys. 50, 859.\\
%
%
Greisen, K. 1966, Phys. Rev. Lett. 16, 748.\\
Hillas, A.M 1971, Proc. of the 12th International
Cosmic Ray Conference, 3, 1001.\\
Hillas, A.M.1981, Proc. of the 17th International
Cosmic Ray Conference, 8, 193.\\
Lawrence, M.A., Reid, R.J. \& Watson, A.A. 1991, J. Phys. G: Nucl. Part. Phys. 17, 733.\\
Migdal, A.B. 1956, Phys. Rev., 103, 1811.\\
Pryke, C.L. 1996, Doctorate Thesis, University of Leeds.\\
Pryke, C.L. 1998, Private Communication.\\
Rhode, W. et al 1996, Nucl. Phys. B (Proc. Suppl.) 49, 491.\\
Simpson, K.M. \& Dawson, B.R. 1996,  GAP-NOTE-96-044, (Reading: http\protect{://}www\protect{.}auger\protect{.}org\protect{/}.)\\
Sokolski, P. et al 1992, Phys. Rep. 217, 5, 226.\\
Zatsepin, G.T. \& Kuzmin, V.A. 1966, JETP Let. 4, 78.\\
\end{document}